\documentclass[journal]{new-aiaa}
\usepackage[utf8]{inputenc}
\usepackage{graphicx}
\usepackage{amsmath}
\usepackage[version=4]{mhchem}
\usepackage{siunitx}
\usepackage{longtable,tabularx}
\usepackage{subcaption}
\usepackage{algorithm}  
\usepackage{algorithmicx}  
\usepackage{algpseudocode}  
\usepackage{float}  
\usepackage{booktabs} 
\usepackage{hyperref}  

\title{Learning-Based Stable Optimal Control for Infinite-Time Nonlinear Regulation Problems}

\author{
Han Wang \footnote{Graduate, School of Astronautics, kingham@buaa.edu.cn.}
and
Di Wu \footnote{Associate Professor, School of Astronautics, wudi2025@buaa.edu.cn}
and
Lin Cheng \footnote{\textbf{Corresponding Author}, Associate Professor, School of Astronautics, chenglin5580@buaa.edu.cn. Member AIAA.}
and
Shengping Gong \footnote{Professor, School of Astronautics, gongsp@buaa.edu.cn. Member AIAA.}
and
Xu Huang \footnote{Associate Professor, School of Astronautics, xunudt@126.com.}
}
\affil{School of Astronautics, Beihang University, Beijing, 102206, China}
\affil{State Key Laboratory of High-Efficiency Reusable Aerospace Transportation Technology, Beijing, 102206, China}
\begin{document}

\maketitle

\begin{abstract}
    Infinite-time nonlinear optimal regulation control is widely utilized in aerospace engineering as a systematic method for synthesizing stable controllers. However, conventional methods often rely on linearization hypothesis, while recent learning-based approaches rarely consider stability guarantees. This paper proposes a learning-based framework to learn a stable optimal controller for nonlinear optimal regulation problems. First, leveraging the equivalence between Pontryagin Maximum Principle (PMP) and Hamilton-Jacobi-Bellman (HJB) equation, we improve the backward generation of optimal examples (BGOE) method for infinite-time optimal regulation problems. A state-transition-matrix-guided data generation method is then proposed to efficiently generate a complete dataset that covers the desired state space. Finally, we incorporate the Lyapunov stability condition into the learning framework, ensuring the stability of the learned optimal policy by jointly learning the optimal value function and control policy. Simulations on three nonlinear optimal regulation problems show that the learned optimal policy achieves near-optimal regulation control and the code is provided at \href{https://github.com/wong-han/PaperNORC}{https://github.com/wong-han/PaperNORC}.
\end{abstract}

\section{Introduction}
The regulation control problem constitutes a fundamental area in the control theory, with broad applications in aerospace engineering, including aircraft altitude cruising \cite{wangRobustNonlinearControl2000,leeAnalysisMissileLongitudinal2019, quExperienceReplayEnhances2025}, satellite attitude orientation\cite{carringtonOptimalNonlinearFeedback1986, liAdaptiveAttitudeStabilization2016, sharifiNonlinearOptimalApproach2020} and quadrotor attitude control\cite{panAttitudeControlQuadrotor2023}. It is essential for these missions to maintain stability and achieve specified optimal performance during the regulation process, which means to solve a nonlinear optimal regulation control problem. Typically, dynamic programming theory formulates its sufficient conditions by constructing the Hamilton-Jacobi-Bellman (HJB) equation \cite{doi:10.1126/science.153.3731.34}. However, it's not trivial to analytically solve this partial differential equation due to the inherent nonlinearity of most dynamic systems \cite{cimenSurveyStateDependentRiccati2012}. To address this problem, this paper aims to propose a learning-based framework to obtain a near-optimal and stable-guaranteed policy for nonlinear optimal regulation control.

For nonlinear regulation control problems, Lyapunov stability theory provides a unified approach to designing a stability-guaranteed controller \cite{sastryLyapunovStabilityTheory1999}. By constructing a positive-definite function, called Lyapunov Control Function (CLF), a stable controller is devised such that the time-derivative of CLF is negative-definite. Lyapunov-based methods have been widely investigated in longitudinal control of hypersonic vehicles \cite{merzMissileAttitudeStabilization2012,anMultipleLyapunovFunctionbased2021} and spacecraft attitude control \cite{chenRobustAttitudeControl2014,liAdaptiveAttitudeStabilization2016,wuFinitetimeAttitudeStabilization2022}. Based on a similar principle, the concept of Barrier Lyapunov Function (BLF) was proposed to address the requirement of state or control constraints \cite{teeBarrierLyapunovFunctions2009, xuBarrierLyapunovFunction2019}. However, the performance of the controller depends on the specific Lyapunov function, and there is no consideration of the optimality of the regulation process.

Optimal control is to address performance optimization while achieving the control objective. The two primary solution conditions for optimal control problems are the PMP and the HJB equation \cite{brysonAppliedOptimalControl1975}. The equivalence between these two conditions are investigated, and they can be transformed into one another under certain scenarios \cite{vinterCostateVariableState1986,zhangUnderstandingConnectionPMP2025}. For classical linear quadratic regulation (LQR) problems, the HJB equation yields an analytical solution of the optimal control policy, which is employed in attitude control and trajectory tracking \cite{doi:10.2514/2.4723, zhangOptimalGuidanceLaw2023}. However, the assumptions of linear dynamics and quadratic cost functions limit its practical applicability. Consequently, the state-dependent Riccati equation (SDRE) approach was proposed to provide approximate solutions to nonlinear optimal regulation problems \cite{PEARSON01111962,linControlClassSecondOrder2018, leeAnalysisMissileLongitudinal2019,chodnickiFinitetimeSDREControl2022}. The nonuniqueness of the state-dependent linear quadratic structure may lead to suboptimal control policies \cite{cimenSurveyStateDependentRiccati2012}. For nonquadratic and nonaffine optimal regulation problems, reference \cite{cimenGlobalOptimalFeedback2004} proposed an approximate sequence of Riccati equation (ASRE) method to obtain the approximate optimal solution. It transforms the nonlinear optimal regulation problem into a sequence of linear quadratic structures, but faces challenges in real-time implementation.

Besides transforming the original problem into a linear quadratic structure, alternative methods approximate the optimal value function to solve the HJB equation. Reference \cite{sznaierRecedingHorizonControl2000} utilizes the SDRE-based value function to approximate the terminal cost function and converts the infinite-time optimal regulation problem into a finite-time problem. However, it also faces difficulties in real-time computation. For numerical approximation, the HJB equation is expanded into Taylor series in \cite{chenInfinitetimeNonlinearQuadratic2004}, while \cite{sharifiNonlinearOptimalApproach2020} employs the Galerkin method, and coefficients can be solved. These methods face a trade-off between approximation order and computational complexity. Recently, learning-based approaches have gained much attention due to the powerful approximation capabilities of neural networks. Approximate dynamic programming (ADP) iteratively learns the optimal value function which is approximated by neural networks \cite{wangRecentProgressReinforcement2024}. Due to the ability of learning optimal policy online, it was increasingly applied to guidance and control in aerospace systems \cite{yangADPBasedSpacecraftAttitude2022, wangAdaptiveFaulttolerantControl2023, wangAdaptiveFaulttolerantOptimal2024, wangRobustIncrementalLearning2025}. In contrast to iterative learning online, supervised learning directly approximates the optimal policy offline, which has higher learning efficiency and avoids the instability of online iteration. In \cite{izzoRealTimeGuidanceLowThrust2021}, a backward generation of optimal examples (BGOE) method was proposed to quickly generate optimal data leveraging PMP conditions, which is applied to finite-time optimal control problems \cite{wangNonlinearOptimalGuidance2022, chengNeuralNetworkBasedNonlinearOptimal2024}. However, it can not be applied to infinite-time optimal regulation problems. Besides, it lacks appropriate initialization rules and the generated data can not cover the desired state space, which is detrimental to neural network training \cite{zhengModelIncrementalLearning2025}.

In conclusion, it remains challenging for infinite-time nonlinear optimal regulation problems to efficiently obtain a near-optimal and stable controller. Inspired by \cite{izzoRealTimeGuidanceLowThrust2021} and \cite{izzoOptimalityPrinciplesSpacecraft2024}, in this paper, we propose a supervised-learning-based framework for nonlinear optimal regulation control and verify its effectiveness on three examples, including a second order nonlinear system, Winged-Cone cruise control and rigid body attitude stabilization. The main contributions can be summarized in three aspects:

1) Leveraging the intrinsic equivalence between the Pontryagin Maximum Principle and the Hamilton-Jacobi-Bellman equation, we extend the BGOE method to infinite-time regulation problems. Specifically, we introduce approximate linear quadratic solution of HJB equation to determine the terminal costate of optimal conditions in the BGOE method. Therefore, the infinite-time regulation problem is transformed into a finite time optimal control problem and the initialization of the backward integration can be settled.

2) To efficiently generate a dataset that covers a desired state space, we propose a state-transition-matrix-guided dataset generation method called STM-BGOE. Compared with the standard BGOE method, STM-BGOE can generate a space-specified dataset, which facilitates the learning process of the neural network.

3) A Lyapunov-based learning framework is proposed to learn both the optimal value function and the optimal control policy. The structure of neural network is designed considering the Lyapunov stability condition such that the stability can be enhanced.

The remainder of this paper is structured as follows. Secion \ref{sec:PFP} describes the infinite-time optimal regulation problem, introduce the BGOE method, and present the motivation of this paper. Section \ref{sec:Method} introduces the proposed learning framework for optimal regulation control. In Section \ref{sec:Simulation}, simulations on three instances are conducted to verify the effectiveness of the proposed learning framework. Section \ref{sec:Conclusion} concludes the paper.

\section{Problem Formulation and Preliminaries}
\label{sec:PFP}
In this study, we focus on infinite-time optimal regulation control for nonlinear systems. The nonlinear optimal regulation problem and its optimal conditions is formulated. To facilitate the following derivation, we introduce the BGOE method for finite-time optimal control, and present the motivation of this paper.

\subsection{Infinite-Time Optimal Regulation Problem}
A continuous-time and time-invariant nonlinear dynamic system can be described by the following equation
\begin{equation}
    \label{system}
    \dot{\boldsymbol{x}}=\boldsymbol{f}\left( \boldsymbol{x},\boldsymbol{u} \right) 
\end{equation}
where $\boldsymbol{x}\in \mathbf{R}^n$ is the state vector, $\boldsymbol{u}\in \mathbf{R}^m$ is the control input vector. For a regulation control problem, the system is assumed satisfying $\dot{\boldsymbol{x}}=0$ at an equilibrium point $(\boldsymbol{x_e}, \boldsymbol{u_e})$. Without lack of generality, we set $\boldsymbol{x_e}=\boldsymbol{0},\boldsymbol{u_e}=\boldsymbol{0}$, and other nonzero equilibrium can be analogized by coordinate transformation.

Define the infinite-time performance index as
\begin{equation}
J\left( \boldsymbol{x}\left( t \right) \right) =\int_t^{\infty}{r\left( \boldsymbol{x}\left( \tau \right) ,\boldsymbol{u}\left( \tau \right) \right) \text{d}\tau}
\end{equation}
where $r\left( \boldsymbol{x,u} \right)$ is the cost function set as a positive-definite function.

The optimal regulation problem is to determine an optimal control policy $\boldsymbol{u}^*(t)$ that lead to the minimal performance index $J^*(\boldsymbol{x}(t))$, i.e.,
\begin{equation}
J^*\left( \boldsymbol{x}\left( t \right) \right) =\underset{u}{\min}\int_t^{\infty}{r\left( \boldsymbol{x}\left( \tau \right) ,\boldsymbol{u}\left( \tau \right) \right) \text{d}\tau}\ =\int_t^{\infty}{r\left( \boldsymbol{x}\left( \tau \right) ,\boldsymbol{u}^*\left( \tau \right) \right) \text{d}\tau}
\end{equation} 

According to the dynamic programming theory, the HJB equation gives the sufficient condition for nonlinear optimal regulation control, which can be expressed as
\begin{equation}
-\frac{\partial J^*\left( \boldsymbol{x}\left( t \right) ,t \right)}{\partial t}=\underset{\boldsymbol{u}\left( t \right)}{\min}\ H\left( \boldsymbol{x}\left( t \right) ,\boldsymbol{u}\left( t \right) ,t \right) =\underset{\boldsymbol{u}\left( t \right)}{\min}\left\{ r\left( \boldsymbol{x,u} \right) +\left[ \frac{\partial J^*\left( \boldsymbol{x}\left( t \right) ,t \right)}{\partial \boldsymbol{x}\left( t \right)} \right] ^T\boldsymbol{f}\left( \boldsymbol{x,u,}t \right) \right\} 
\end{equation}
where $H\left( \boldsymbol{x},\boldsymbol{u},t \right)$ is the Hamiltonian function. 

Since the system is assumed to be time-invariant, the HJB equation can be rewritten as
\begin{equation}
	\label{HJB}
r\left( \boldsymbol{x,u}^* \right) +\left[ \frac{\text{d}J^*\left( \boldsymbol{x} \right)}{\text{d}\boldsymbol{x}} \right] ^T\boldsymbol{f}\left( \boldsymbol{x,u}^* \right) =0
\end{equation}

Due to the nonlinearity of the dynamic system, most of the time it's intractable to analytically solve the HJB equation \cite{wangAdaptiveDynamicProgramming2009}. Therefore, researchers resort to approximate solution methods such as ADP methods, where a neural network is used to approximate the optimal value function \cite{wangRecentProgressReinforcement2024}. However, the iterative training process is computationally slow and the derived control policy may be unstable due to approximation errors. Instead of iteratively solving the approximate optimal value function like ADP methods, we propose a supervised learning framework by extending the BGOE method to infinite-time optimal regulation problems. The BGOE method, leveraging the PMP condition, can efficiently generate optimal data for optimal control problems. Before we formally present the proposed learning framework, a brief introduction of the BGOE method is first presented in the next subsection.

\subsection{Introduction to the Back Generation of Optimal Examples}
The main idea of BGOE method, initially proposed by Izzo et al. \cite{izzoRealTimeGuidanceLowThrust2021} is to rapidly generate optimal data by performing backward integration of the Hamiltonian system, which is derived by the minimum principle. It is typically used to solve finite-time optimal control problems with terminal constraints.

For finite-time optimal control problems, the performance index can be expressed as
\begin{equation}
J=\int_0^T{r\left( \boldsymbol{x}\left( t \right) ,\boldsymbol{u}\left( t \right) \right)}\text{d}t
\end{equation}

Optimal control policy should satisfy the following terminal constraint:
\begin{equation}
    \label{BGOE-TerminalConstraint}
\boldsymbol{g}\left( \boldsymbol{x}\left( T \right) ,T \right) =0
\end{equation}

According to the PMP condition, the Hamiltonian can be expressed as 
\begin{equation}
\label{BGOE-Hamiltonian}
H=r\left( \boldsymbol{x}\left( t \right) ,\boldsymbol{u}\left( t \right) \right) +\boldsymbol{p}^T\left( t \right) \boldsymbol{f}\left( \boldsymbol{x}\left( t \right) ,\boldsymbol{u}\left( t \right) \right) 
\end{equation}
where $\boldsymbol{p}$ represents the costate vector, and its time derivative can be derived as
\begin{equation}
    \label{BGOE-p_dot}
\boldsymbol{\dot{p}}=-\frac{\partial H}{\partial \boldsymbol{x}}=-\frac{\partial r\left( \boldsymbol{x,u} \right)}{\boldsymbol{x}}-\left( \frac{\partial \boldsymbol{f}\left( \boldsymbol{x,u} \right)}{\partial \boldsymbol{x}} \right) ^T\boldsymbol{p}
\end{equation}

The traverse condition and the extreme condition can be expressed as
\begin{equation}
  \label{terminal costate}
\boldsymbol{p}\left( T \right) =\frac{\partial \boldsymbol{g}^T\left( \boldsymbol{x}\left( T \right) ,T \right)}{\partial \boldsymbol{x}\left( T \right)}\boldsymbol{\nu }
\end{equation}
\begin{equation}
\label{BGOE-U}
\boldsymbol{u}^*=\underset{\boldsymbol{u}\left( t \right) \in U}{\min}H\left( \boldsymbol{x}^*\left( t \right) ,\boldsymbol{u}\left( t \right) ,\boldsymbol{p}\left( t \right) \right) 
\end{equation}

Equation (\ref{BGOE-TerminalConstraint})-(\ref{BGOE-U}) constitute the optimal condition for the finite-time optimal control problem. The state and costate constitute the Hamiltonian system as follows
\begin{equation}
  \label{BGOE-HamiltonianSystem}
\left\{ \begin{aligned}
	\boldsymbol{\dot{x}}&=-\frac{\partial H}{\partial \boldsymbol{p}}\\
	\boldsymbol{\dot{p}}&=-\frac{\partial H}{\partial \boldsymbol{x}}\\
\end{aligned} \right. 
\end{equation}

Given a specific terminal state $\boldsymbol{x}(T)$ and terminal costate $\boldsymbol{p}(T)$ satisfying equation (\ref{BGOE-TerminalConstraint}) and (\ref{terminal costate}) respectively, the BGOE method generates optimal trajectories satisfying above optimal conditions by backward integrating the Hamiltonian system (\ref{BGOE-HamiltonianSystem}) from the terminal time to the initial time.

The BGOE method provide an effective method for fast generating optimal data for finite-time optimal control problems. However, it can not be applied to infinite-time optimal regulation problems because the backward integration process can not start from the equilibrium and the integration time can not be set to infinity. Besides, it does not provide a selection rule for the terminal costate in the standard BGOE method. The generated optimal trajectory is non-deterministic, causing the generated dataset can not cover the desired state space, limiting the applicability of the learned policy. To address these two drawbacks, we will provide our method in the next subsection.

\section{Learning Stable Nonlinear Optimal Regulation Control}
\label{sec:Method}
To efficiently generate an optimal dataset for nonlinear optimal regulation problems, we first extend the BGOE method to infinite-time regulation control by combining the HJB equation and the PMP condition. Then we propose a state-transition-matrix-guided dataset generation method that can generate a complete dataset that covers the desired state space. To learn a stable optimal control policy from the generated dataset, we propose a supervised-learning framework where the Lyapunov stability condition is integrated. 

\subsection{Extending BGOE for Infinite-Time Optimal Regulation Control}
To extend the BGOE method to infinite-time optimal regulation control problems, we first transform the infinite-time control problem into a finite-time problem with an optimal terminal cost. The terminal cost is approximated by the approximate solution of the HJB equation around the equilibrium point. According to the equivalence between the HJB equation and PMP condition, the optimal trajectory of the original nonlinear problem can be obtained leveraging the BGOE method.

According to the Bellman optimality condition, the infinite-time performance index can be rewritten as
\begin{equation}
  \label{finite-time J}
\begin{aligned}
	J^*\left( \boldsymbol{x}\left( t \right) \right) &=\underset{\boldsymbol{u}}{\min}\left\{ \int_t^T{r\left( \boldsymbol{x}\left( \tau \right) ,\boldsymbol{u}\left( \tau \right) \right) \text{d}\tau}+\int_T^{\infty}{r\left( \boldsymbol{x}\left( \tau \right) ,\boldsymbol{u}\left( \tau \right) \right) \text{d}\tau} \right\}\\
	&=\underset{\boldsymbol{u}}{\min}\left\{ \int_t^T{r\left( \boldsymbol{x}\left( \tau \right) ,\boldsymbol{u}\left( \tau \right) \right) \text{d}\tau +J^*\left( \boldsymbol{x}\left( T \right) \right)} \right\}\\
\end{aligned}
\end{equation}

Therefore, the optimal solution for the infinite-time problem can be obtained by separately solving the optimal terminal cost $J^*\left( \boldsymbol{x}\left( T \right) \right)$ and a finite-time optimal control problem. We first derive the optimal terminal cost by approximating the HJB equation around the equilibrium point. The linear approximation of the nonlinear system and the performance index can be expressed as
\begin{equation}
\boldsymbol{\dot{x}}=\boldsymbol{Ax}+\boldsymbol{Bu}
\end{equation}
\begin{equation}
\hat{J}\left( \boldsymbol{x}(T) \right) =\int_T^{\infty}{\boldsymbol{x}^T\boldsymbol{Qx}+\boldsymbol{u}^T\boldsymbol{Ru}\text{d}\tau}
\end{equation}
where 
\begin{equation}
\boldsymbol{A}=\left. \frac{\partial \boldsymbol{f}\left( \boldsymbol{x,u} \right)}{\partial \boldsymbol{x}} \right|_{\boldsymbol{x}_{\boldsymbol{e}},\boldsymbol{u}_{\boldsymbol{e}}},\boldsymbol{B}=\left. \frac{\partial \boldsymbol{f}\left( \boldsymbol{x,u} \right)}{\partial \boldsymbol{x}} \right|_{\boldsymbol{x}_{\boldsymbol{e}},\boldsymbol{u}_{\boldsymbol{e}}},\boldsymbol{Q}=\left. \frac{\partial ^2r\left( \boldsymbol{x,u} \right)}{\partial \boldsymbol{x}^2} \right|_{\boldsymbol{x}_{\boldsymbol{e}},\boldsymbol{u}_{\boldsymbol{e}}},\boldsymbol{R}=\left. \frac{\partial ^2r\left( \boldsymbol{x,u} \right)}{\partial \boldsymbol{u}^2} \right|_{\boldsymbol{x}_{\boldsymbol{e}},\boldsymbol{u}_{\boldsymbol{e}}}
\end{equation}

The approximate problem forms a typical linear quadratic regulation problem. The optimal terminal cost can be approximated by the optimal value function of the linear quadratic regulation problem, which is expressed as
\begin{equation}
\hat{J}^*\left( \boldsymbol{x}(T) \right) =\boldsymbol{x}^T\boldsymbol{Px}
\end{equation}
where $\boldsymbol{P}$ is the non-negative definite solution of the Riccati equation
\begin{equation}
\boldsymbol{PA}+\boldsymbol{A}^T\boldsymbol{P}-\boldsymbol{PBR}^{-1}\boldsymbol{B}^T\boldsymbol{P}+\boldsymbol{Q}=0
\end{equation}

Note that the linear quadratic approximate solution is only locally accurate around the equilibrium point. However, we can obtain the optimal solution of the original nonlinear problem by solving the finite-time optimal control problem (\ref{finite-time J}) leveraging the BGOE method.

Define the Hamiltonian function and the corresponding Hamiltonian system
\begin{equation}
H\left( \boldsymbol{x,p,u} \right) =r\left( \boldsymbol{x,u} \right) +\boldsymbol{p}^T\boldsymbol{f}\left( \boldsymbol{x,u} \right) 
\end{equation}
\begin{equation}
\left\{ \begin{aligned}
    \label{NOR-BGOE-Hamiltonian}
	\boldsymbol{\dot{x}}&=-\frac{\partial H}{\partial \boldsymbol{p}}=\boldsymbol{f}\left( \boldsymbol{x,u} \right)\\
	\boldsymbol{\dot{p}}&=-\frac{\partial H}{\partial \boldsymbol{x}}=\frac{\partial r\left( \boldsymbol{x,u} \right)}{\partial \boldsymbol{x}}+\left[ \frac{\partial \boldsymbol{f}\left( \boldsymbol{x,u} \right)}{\partial \boldsymbol{x}} \right] ^T\boldsymbol{p}\\
\end{aligned} \right. 
\end{equation}

The extreme condition can be defined through the PMP condition, which is given as
\begin{equation}
\boldsymbol{u}^*=\underset{\boldsymbol{u}\left( t \right) \in U}{\min}H=\underset{\boldsymbol{u}\left( t \right) \in U}{\min}r\left( \boldsymbol{x,u} \right) +\boldsymbol{p}^T\boldsymbol{f}\left( \boldsymbol{x,u} \right) 
\end{equation}

Leveraging the equivalence between the HJB equation and the PMP condition \cite{brysonAppliedOptimalControl1975,vinterCostateVariableState1986}, we can determine the traverse condition for the finite-time optimal control problem, which is expressed as
\begin{equation}
	\label{boundary condition}
\left\{ \begin{aligned}
	\boldsymbol{x}( t ) &=\boldsymbol{x}_0\\
	\boldsymbol{x}( T ) &=\boldsymbol{x}_f\\
	\boldsymbol{p}( T ) &=\frac{\partial \hat{J}^*\left( \boldsymbol{x}\left( T \right) \right)}{\partial \boldsymbol{x}\left( T \right)}=2\boldsymbol{Px}_f\\
\end{aligned} \right. 
\end{equation}

Note that the terminal state $\boldsymbol{x}_f$ can be chosen from the punctured neighborhood of the equilibrium point such that the linearized approximation is valid.
\begin{equation}
    \label{Neighborhood}
\boldsymbol{x}_f\in \mathcal{B},\,\,\mathcal{B}=\left\{ \left. \boldsymbol{x} \right|\lVert \boldsymbol{x}-\boldsymbol{x}_{\boldsymbol{e}} \rVert _2\le \delta ,\,\,\boldsymbol{x}\ne \boldsymbol{x}_{\boldsymbol{e}} \right\} 
\end{equation}

Therefore, we can initialize the BGOE method for infinite-time optimal regulation problems. According to optimal conditions (\ref{NOR-BGOE-Hamiltonian})-(\ref{Neighborhood}), one can backward integrate the Hamiltonian system from any terminal state in region $\mathcal B$ and the corresponding optimal trajectory can be easily obtained. 

The procedure for generating optimal trajectories for infinite-time regulation problems is concluded in Algorithm \ref{Algorithm1}. Next, we will further introduce the state-transition-matrix-guided dataset generation method, which provides a formal rule to choose the terminal state $\boldsymbol{x}_f$ and integration time $T$, such that a complete dataset can be generated. 

\floatname{algorithm}{Algorithm}
\begin{algorithm*}[!h]
	
  \caption{\label{Algorithm1} BGOE for infinite-time optimal regulation control}
  \begin{algorithmic}[1]
	\Statex \textbf{Input:} terminal state $\boldsymbol{x}_f$, terminal time $T$

	\Statex \textbf{Output:} Optimal trajectory data $\mathcal{D}=\left\{ \boldsymbol{x}^*, \boldsymbol{u}^*, J^* \right\}$

	\State Initialize: Set $\mathcal D=\left\{\right\}$, $t=T$, $\boldsymbol{x}^*(t)=\boldsymbol{x}_f$, $\boldsymbol{p}^*(t)=2\boldsymbol{P}\boldsymbol{x}_f$ and $J^*(\boldsymbol{x}_f)=\boldsymbol{x}_f^T\boldsymbol{P}\boldsymbol{x}_f$.
	\For {$t$ from $T$ to $0$}
		\State Backward integrate the Hamiltonian system (\ref{NOR-BGOE-Hamiltonian}) to obtain $(\boldsymbol{x}^*, \boldsymbol{u}^*)$.
        \State Accumulate performance index $J^*(\boldsymbol{x}(t))=\int{r\left( \boldsymbol{x}^*(t),\boldsymbol{u}^*(t) \right) \text{d}t} + J^*(\boldsymbol{x}_f)$.
        \State Store $(\boldsymbol{x}^*, \boldsymbol{u}^*, J^*)$ in $\mathcal D$.
	\EndFor

	\State End the algorithm. Store optimal dataset $\mathcal D$.
  \end{algorithmic}
\end{algorithm*}

\subsection{State-Transition-Matrix-Guided Specific Data Generation}
The standard BGOE method lacks a selection rule for terminal states. Selecting terminal states randomly or uniformly can not guarantee the generated dataset covers the desired state space. Therefore, by calculating the state transition matrix of Hamiltonian system, we derive a selection rule to efficiently generate a complete dataset that covers the desired state space, which facilitates the learning of control policy.  

To determine how to set the terminal state $\boldsymbol{x_f}$ and terminal time $T$ for generating the desired dataset, we first analyze the relationship between these parameters and the generated state. Define the augmented state of Hamiltonian system as $\boldsymbol{Z}=\left[ \boldsymbol{x}^T,\boldsymbol{p}^T \right] ^T$. Then we can rewrite the Hamiltonian system as 

\begin{equation}
    \label{parameterized system}
\boldsymbol{\dot{Z}}=\left[ \begin{aligned}
	\boldsymbol{\dot{x}}\\
	\boldsymbol{\dot{p}}\\
\end{aligned} \right] =\left[ \begin{aligned}
	-\frac{\partial H}{\partial \boldsymbol{p}}\\
	-\frac{\partial H}{\partial \boldsymbol{x}}\\
\end{aligned} \right] =\boldsymbol{F}\left( \boldsymbol{Z} \right) 
\end{equation}

The state transition matrix of the Hamiltonian system can be expressed as
\begin{equation}
    \label{STM}
\delta \boldsymbol{Z}\left( t \right) =\boldsymbol{\Phi }\left( T,t \right) \delta \boldsymbol{Z}\left( T \right) , s.t. \boldsymbol{\Phi }\left( T,T \right) =\boldsymbol{I}
\end{equation}
which characterizes the relationship between the change of terminal state $\delta \boldsymbol{Z}\left( T \right)$ and the corresponding change $\delta \boldsymbol{Z}\left( t \right)$ after a backward integration from $T$ to $t$. 

The time derivative of the state transition matrix can be derived by chain rule as
\begin{equation}
	\label{STM DF}
\begin{aligned}
	\boldsymbol{\dot{\Phi}}\left( T,t \right) &=\frac{\text{d}}{\text{d}t}\left( \frac{\partial \boldsymbol{Z}\left( t \right)}{\partial \boldsymbol{Z}\left( T \right)} \right) =\frac{\partial}{\partial \boldsymbol{Z}\left( T \right)}\left( \frac{\text{d}\boldsymbol{Z}\left( t \right)}{\text{d}t} \right)\\
	&=\frac{\text{d}\boldsymbol{F}\left( \boldsymbol{Z}\left( t \right) \right)}{\text{d}\boldsymbol{Z}\left( t \right)}\frac{\partial \boldsymbol{Z}\left( t \right)}{\partial \boldsymbol{Z}\left( T \right)}\\
	&=\boldsymbol{F}_{\boldsymbol{Z}}\boldsymbol{\Phi }\left( T,t \right)\\
\end{aligned}
\end{equation}
where $\boldsymbol{F}_{\boldsymbol{Z}}$ denotes the Jacobian matrix of the Hamiltonian system.

By simultaneously backward integrating the Hamiltonian system (\ref{parameterized system}) and the equation (\ref{STM DF}) from $T$ to $t_0$, we can obtain the state transition matrix of the generated optimal trajectory, denoted as $\boldsymbol{\Phi }\left( T,t_0 \right)$. 

Note that $\boldsymbol{p}(T)=2\boldsymbol{Px}_f$ in equation (\ref{boundary condition}), when there are small variations of terminal state $\delta \boldsymbol{x}_f$ and integration time $\delta t$, the corresponding change of generated state can be derived as follows
\begin{equation}
	\label{relationship}
\begin{aligned}
	\delta \boldsymbol{x}\left( t_0 \right) &=\big[\boldsymbol{I} \quad \boldsymbol{0}\big] \delta \boldsymbol{Z}\left( t_0 \right)\\
	&=\big[\boldsymbol{I} \quad \boldsymbol{0}\big] \left[ \boldsymbol{\Phi }\left( T,t_0+\delta t \right) \delta \boldsymbol{Z}\left( T \right) +\frac{\text{d}\boldsymbol{Z}\left( t_0 \right)}{\text{d}t}\delta t \right]\\
	&=\big[\boldsymbol{I} \quad \boldsymbol{0}\big] \boldsymbol{\Phi }\left( T,t_0+\delta t \right) \left[ \begin{array}{c}
	\delta \boldsymbol{x}\left( T \right)\\
	\delta \boldsymbol{p}\left( T \right)\\
\end{array} \right] +\frac{\text{d}\boldsymbol{x}\left( t_0 \right)}{\text{d}t}\delta t\\
	&=\big[\boldsymbol{I} \quad \boldsymbol{0}\big] \left( \boldsymbol{\Phi }\left( T,t_0 \right) +\boldsymbol{\dot{\Phi}}\left( T,t_0 \right) \delta t \right) \left[ \begin{array}{c}
	\boldsymbol{I}\\
	2\boldsymbol{P}\\
\end{array} \right] \delta \boldsymbol{x}_f+\frac{\text{d}\boldsymbol{x}\left( t_0 \right)}{\text{d}t}\delta t\\
	&\approx \big[\boldsymbol{I} \quad \boldsymbol{0}\big] \boldsymbol{\Phi }\left( T,t_0 \right) \left[ \begin{array}{c}
	\boldsymbol{I}\\
	2\boldsymbol{P}\\
\end{array} \right] \delta \boldsymbol{x}_f+\frac{\text{d}\boldsymbol{x}\left( t_0 \right)}{\text{d}t}\delta t\\
	&=\frac{\partial \boldsymbol{x}\left( t_0 \right)}{\partial \boldsymbol{x}\left( T \right)}\delta \boldsymbol{x}_f+\frac{\text{d}\boldsymbol{x}\left( t_0 \right)}{\text{d}t}\delta t\\
\end{aligned}
\end{equation}

Based on the above equation, we can manipulate the generated optimal data $\boldsymbol x(t_0)$ by changing the selected terminal state $\boldsymbol x_f$ and integration time $T$. Above equation is underdetermined and has multiple solutions. One can transform it into a constrained minimization problem as
\begin{equation}
	\label{min adjust}
\begin{aligned}
	&\underset{\delta \boldsymbol{x}_f,\delta t}{\min}\ \lVert \delta \boldsymbol{x}_f \rVert _2\\
	&s.t. \ \delta \boldsymbol{x}\left( t_0 \right) =\frac{\partial \boldsymbol{x}\left( t_0 \right)}{\partial \boldsymbol{x}\left( T \right)}\delta \boldsymbol{x}_f+\frac{\text{d}\boldsymbol{x}\left( t_0 \right)}{\text{d}t}\delta t\\
  &\boldsymbol{x}_f \in \mathcal{B}
\end{aligned}
\end{equation}

When $\delta \boldsymbol{x}\left( t_0 \right)$ is small enough, we can set $\delta t=0$ and the $\delta \boldsymbol{x}_f$ can be directly calculated as
\begin{equation}
	\label{simple adjust}
\delta \boldsymbol{x}_f=\left[ \frac{\partial \boldsymbol{x}\left( t_0 \right)}{\partial \boldsymbol{x}\left( T \right)} \right] ^{-1}\delta \boldsymbol{x}\left( t_0 \right)
\end{equation}

According to the equation (\ref{min adjust}) or equation (\ref{simple adjust}), we can determine the input parameter in Algorithm \ref{Algorithm1}. Therefore, we provide a state-transition-matrix data generation method, called STM-BGOE, to consecutively generate optimal data that covers the desired state space. The algorithm is summarized in Algorithm \ref{Algorithm2}. For the legibility of the pseudo-code, we omit the details of setting desired initial states $\boldsymbol{x}_0^d$ in step \ref{loop}. It is sequentially chosen from the desired region to make sure $\delta \boldsymbol{x}_0$ is small enough.

\floatname{algorithm}{Algorithm}
\begin{algorithm*}[!h]
	
  \caption{\label{Algorithm2} State-transition-matrix-guided backward generation of optimal examples}
  \begin{algorithmic}[1]
	\Statex \textbf{Input:} Desired region of initial states $\boldsymbol{x}_0\in \mathcal D_d=\left\{ \boldsymbol{x}_0\in \mathbf{R}^n|a_1\le x_1\le b_1,a_2\le x_2\le b_2,\cdots ,a_n\le x_n\le b_n \right\}$

	\Statex \textbf{Output:} Optimal dataset whose initial states cover the desired state region $\mathcal{D}=\left\{ \boldsymbol{x}^*, \boldsymbol{u}^*, J^* \right\}$

	\State Initialize: Set $\mathcal D=\left\{\right\}$, terminal state $\boldsymbol{x}_f$, terminal time $T$, state transition matrix $\boldsymbol{\Phi } =\boldsymbol{I}$
	\State \label{generation step}Calculate the corresponding optimal dataset $\left\{ \boldsymbol{x}^*, \boldsymbol{u}^*, J^* \right\}$ by Algorithm \ref{Algorithm1} and extract the generated initial state $\boldsymbol{x}^*_0$. Simultaneously calculate the state transition matrix $\boldsymbol{\Phi }\left( T,t_0 \right)$ according to (\ref{STM DF}).

	\For {each desired initial state $\boldsymbol{x}_0^d$ in the desired region} \label{loop}
		\State Calculate $\delta \boldsymbol{x}_0=\boldsymbol{x}_0^d - \boldsymbol{x}_0^*$
		\State Calculate $\delta \boldsymbol{x}_f$ and $\delta t$ according to equation (\ref{min adjust}) or equation (\ref{simple adjust}).
		\State Update the terminal state $\boldsymbol{x}_f$ and terminal time $T$
		\State Implement step \ref{generation step} again to generate the optimal dataset $\left\{ \boldsymbol{x}^*, \boldsymbol{u}^*, J^* \right\}$, the initial state $\boldsymbol{x}^*_0$ and the state transition matrix $\boldsymbol{\Phi }\left( T,t_0 \right)$.
	\EndFor

	\State End the algorithm. Store optimal dataset $\mathcal D$.
  \end{algorithmic}
\end{algorithm*}

\subsection{Lyapunov-Based Optimal Strategy Learning Framework}
Based on the generated optimal data, we design a Lyapunov-based optimal strategy learning framework, where two neural networks are designed to approximate the optimal value function and the optimal control policy, respectively. The Lyapunov stability condition is imposed on the neural networks to guarantee the stability of the learned control policy.

According to the HJB equation (\ref{HJB}), the optimal value function and the optimal control policy inherently satisfy the following Lyapunov stability conditions
\begin{equation}
	\label{Lyapunov condition}
\begin{aligned}
	J\left( \boldsymbol{x} \right) &>0\ \forall \boldsymbol{x}\ne 0,\ J\left( 0 \right) =0\\
	\dot{J}\left( \boldsymbol{x} \right) &=\left[ \frac{\text{d}J^*\left( \boldsymbol{x} \right)}{\text{d}\boldsymbol{x}} \right] ^T\boldsymbol{f}\left( \boldsymbol{x,u}^* \right) =-r\left( \boldsymbol{x,u}^* \right) \le 0\\
\end{aligned}
\end{equation}

Based on this condition, we design a neural network $V_{\text{net}}$ to approximate the optimal value function, which is expressed as follows
\begin{equation}
	\label{V_net}
V_{\text{net}}\left( \boldsymbol{x};\boldsymbol{\theta} \right) =\left( \boldsymbol{x}-\boldsymbol{x}_{\boldsymbol{e}} \right) ^T\boldsymbol{L}\left( \boldsymbol{x};\boldsymbol{\theta} \right) \boldsymbol{L}\left( \boldsymbol{x};\boldsymbol{\theta} \right) ^T\left( \boldsymbol{x}-\boldsymbol{x}_{\boldsymbol{e}} \right) 
\end{equation}
where $\boldsymbol{L}$ is a lower triangular matrix with positive diagonal elements, and $\boldsymbol{\theta}$ is the parameter of the neural network. According to the Cholesky decomposition theorem, $\boldsymbol{L}\left( \boldsymbol{x};\boldsymbol{\theta} \right) \boldsymbol{L}\left( \boldsymbol{x};\boldsymbol{\theta} \right) ^T$ is a positive definite matrix. Therefore, $V_{\text{net}}\left( \boldsymbol{x};\boldsymbol{\theta} \right)$ satisfies the first condition in equation (\ref{Lyapunov condition}).

For the optimal control policy, we design a simple deep neural network $\boldsymbol{u}_{\text{net}}$ to approximate the optimal control command, but we perform a correction on the actual output of the neural network to satisfy the second condition in equation (\ref{Lyapunov condition}). Therefore, the approximate optimal control policy can be expressed as
\begin{equation}
	\label{u_hat}
\boldsymbol{\hat{u}}=\left\{ \begin{matrix}
	\boldsymbol{u}_{\text{net}},&		\dot{V}_{\text{net}}\le 0\\
	\boldsymbol{u}_{\text{net}}+\delta \boldsymbol{u,}&		\dot{V}_{\text{net}}>0\\
\end{matrix} \right. 
\end{equation}
\begin{equation}
  \dot{V}_{\text{net}}=\left( \frac{\partial V_{\text{net}}}{\partial \boldsymbol{x}} \right) ^T \boldsymbol{f}\left( \boldsymbol{x,u}_{\text{net}} \right)
\end{equation}
where the derivative of approximate value function $ \frac{\partial V_{\text{net}}}{\partial \boldsymbol{x}}$ can be obtained by performing a backward propagation through the neural network.

The correction of control command $\delta \boldsymbol{u}$ can be calculated by solving the following minimization problem
\begin{equation}
\begin{aligned}
	&\underset{\delta \boldsymbol{u}}{\min}\ \lVert \delta \boldsymbol{u} \rVert _2\\
	&s.t.\ \left( \frac{\partial V_{\text{net}}}{\partial \boldsymbol{x}} \right) ^T \boldsymbol{f}\left( \boldsymbol{x,u}_{\text{net}}+\delta \boldsymbol{u} \right) \le -k\lVert \boldsymbol{x} \rVert _2\\
\end{aligned}
\end{equation}
where $k$ is a small positive constant to ensure $\dot{V}_{\text{net}}$ is negative definite. 

For nonlinear affine systems $\boldsymbol{\dot{x}}=\boldsymbol{f}_a\left( \boldsymbol{x} \right) +\boldsymbol{g}_a\left( \boldsymbol{x} \right) \boldsymbol{u}$, the optimal solution of above problem can be directly calculated as
\begin{equation}
\delta \boldsymbol{u}=-\boldsymbol{A}^T\left( \boldsymbol{AA}^T \right) ^{-1}\left( \dot{V}_{\text{net}}+k\lVert \boldsymbol{x} \rVert _2 \right)
\end{equation}
\begin{equation}
\boldsymbol{A}=\left( \frac{\partial V}{\partial \boldsymbol{x}} \right) ^T\boldsymbol{g}_a\left( \boldsymbol{x} \right)
\end{equation}

By simultaneously learning the optimal value function and the optimal control policy and imposing the Lyapunov stability condition, the stability of the approximate optimal control command is guaranteed.

\begin{figure*}[hbt!]
  \centering
  \includegraphics[width=0.9\textwidth]{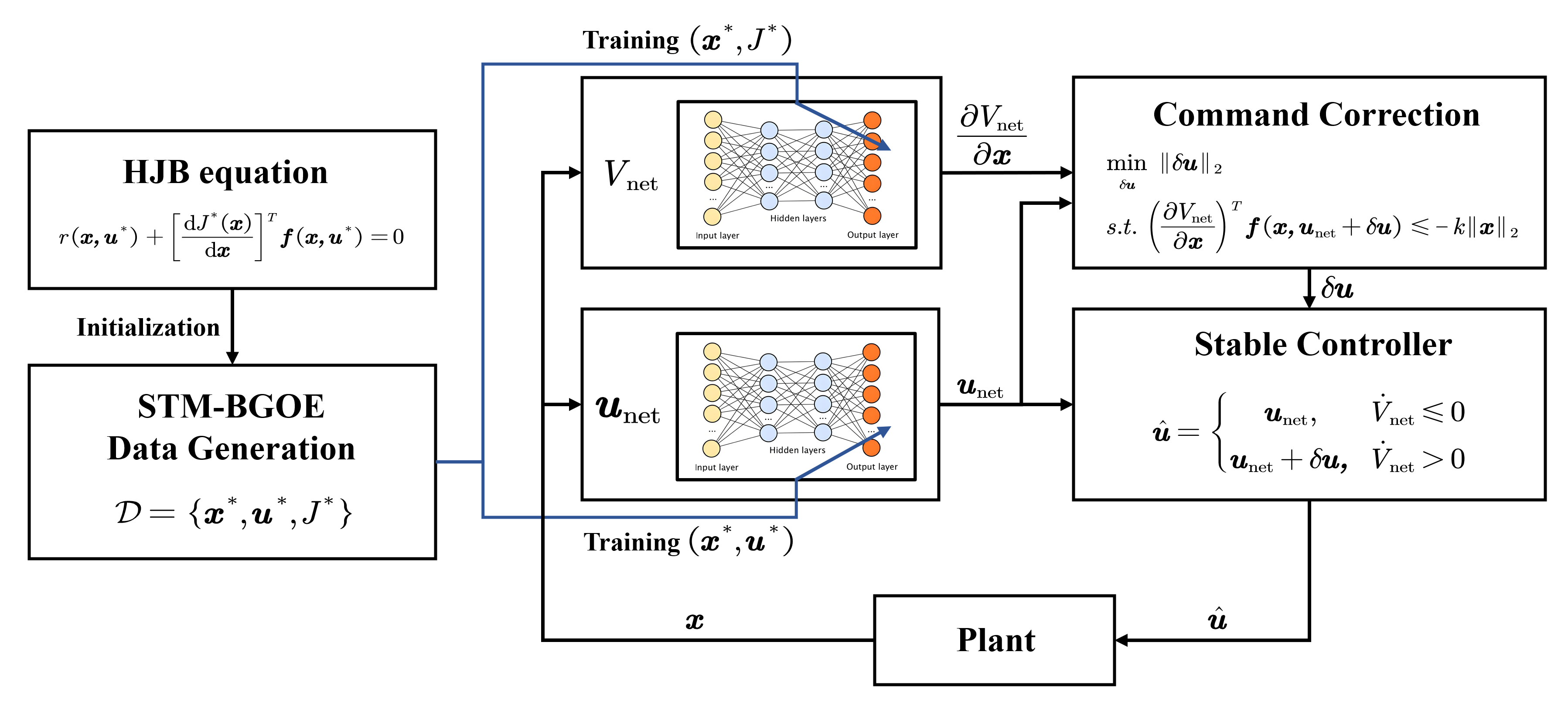}
  \caption{\label{fig:Framework} Framework for learning stable nonlinear optimal regulation control}
\end{figure*}

In summary, in this section, a learning-based framework for stable nonlinear optimal regulation control, illustrated as Fig. \ref{fig:Framework}. Leveraging the proposed STM-BGOE method, an optimal dataset that covers a desired state space can be efficiently generated. The optimal value function and control policy are devised satisfying the Lyapunov stability condition such that the learned control policy is guaranteed to be stable. Next we will conduct simulations on three examples to validate the effectiveness of the proposed method.

\section{Simulation}
\label{sec:Simulation}
To verify the effectiveness of the proposed method, we conduct simulations on three different nonlinear optimal regulation problems, including a second-order nonlinear system with known optimal solution for validation, a Winged-Cone model cruise control problem and a rigid body attitude stabilization problem. Compared with the LQR controller, the proposed method exhibits superior performance index. The simulation is performed using Python and the code can be accessed at \href{https://github.com/wong-han/PaperNORC}{https://github.com/wong-han/PaperNORC}

\subsection{Nonlinear Second Order System}
Considering the following second-order nonlinear system \cite{muDynamicEventTriggeringNeural2022}
\begin{equation}
\left\{ \begin{aligned}
	\dot{x}_1&=-x_1+x_2\\
	\dot{x}_2&=-0.5x_1-0.5x_2\left( 1-\left( \cos \left( 2x_1 \right) +2 \right) ^2 \right) +\left( \cos \left( 2x_1 \right) +2 \right) u\\
\end{aligned} \right. 
\end{equation}

\begin{figure}[!htb]
  \centering
\begin{subfigure}{0.49\textwidth}
  \centering
  \includegraphics[scale=1.0]{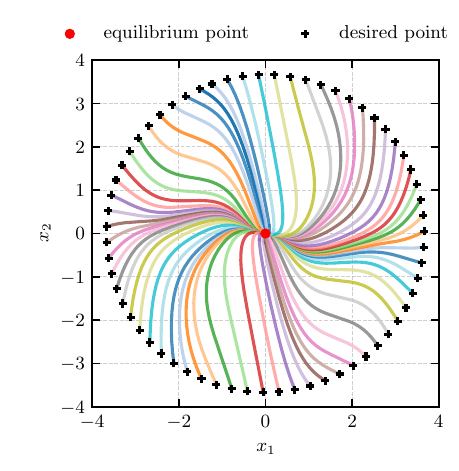}
  \caption{\label{fig:optimal_data_SOP} STM-BGOE}
\end{subfigure}
\hfill
\begin{subfigure}{0.49\textwidth}
  \centering
  \includegraphics[scale=1.0]{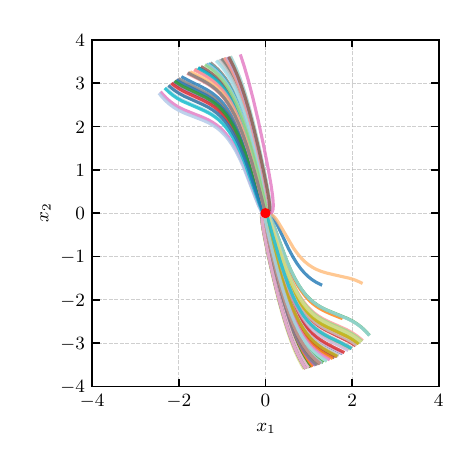}
  \caption{\label{fig:optimal_data_SOP_NoSTM} Standard BGOE}
\end{subfigure}
\caption{Comparison of optimal data distribution}
\label{fig:optimal_data_comparison}
\end{figure}

\begin{figure}[!htb]
  \centering
\begin{subfigure}{0.49\textwidth}
  \centering
  \includegraphics[scale=1.0]{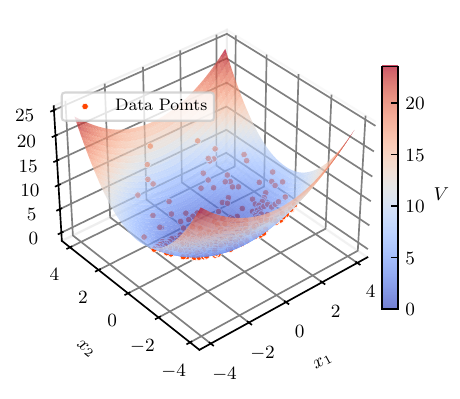}
  \caption{\label{fig:value_function_SOP} Approximate optimal value function}
\end{subfigure}
\hfill
\begin{subfigure}{0.49\textwidth}
  \centering
  \includegraphics[scale=1.0]{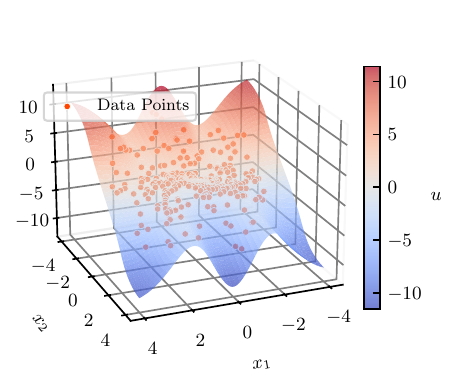}
  \caption{\label{fig:action_SOP} Approximate optimal policy}
\end{subfigure}
\caption{The learned optimal value function and optimal control policy}
\label{fig:optimal_value_policy_sop}
\end{figure}

\begin{figure}[!htb]
  \centering
\begin{subfigure}{0.49\textwidth}
  \centering
  \includegraphics[scale=1.0]{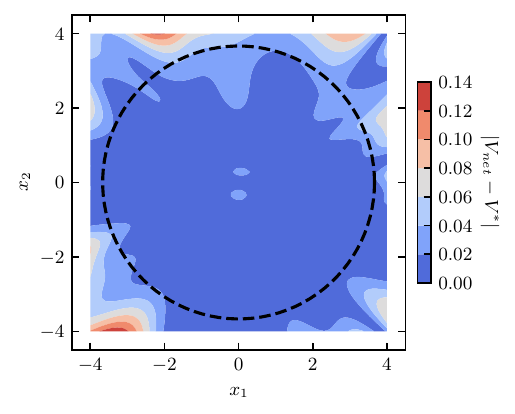}
  \caption{\label{fig:value_function_SOP_Compare} Optimal value function approximation error}
\end{subfigure}
\hfill
\begin{subfigure}{0.49\textwidth}
  \centering
  \includegraphics[scale=1.0]{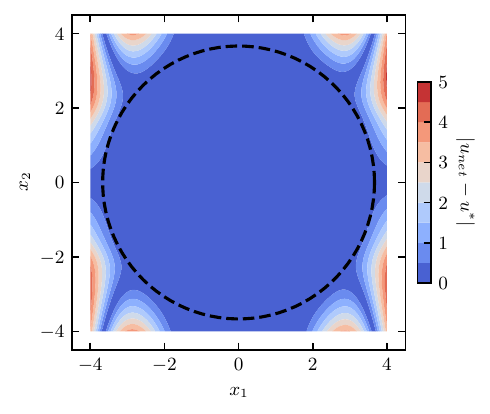}
  \caption{\label{fig:action_SOP_Compare} Optimal policy approximation error}
\end{subfigure}
\caption{The approximation error of optimal value function and optimal control policy}
\label{fig:optimal_value_policy_error_sop}
\end{figure}

The performance index is set as
\begin{equation}
J=\int_0^{\infty}{\boldsymbol{x}^T\boldsymbol{Qx}+Ru^2\text{d}\tau}
\end{equation}
where $\boldsymbol{Q}$ are identity matrices and $R=1$. This problem has an analytical optimal solution which can be used for validation, expressed as
\begin{equation}
\begin{aligned}
	J^*&=\frac{1}{2}x_{1}^{2}+x_{2}^{2}\\
	u^*&=-\left( \cos \left( 2x_1 \right) +2 \right) x_2\\
\end{aligned}
\end{equation}

\begin{figure}[!htb]
  \centering
\begin{subfigure}{0.49\textwidth}
  \centering
  \includegraphics[scale=1.0]{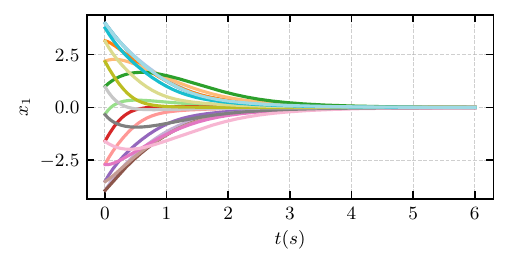}
  \caption{\label{fig:x1_SOP} $x_1$-$t$ profiles}
\end{subfigure}
\vfill
\begin{subfigure}{0.49\textwidth}
  \centering
  \includegraphics[scale=1.0]{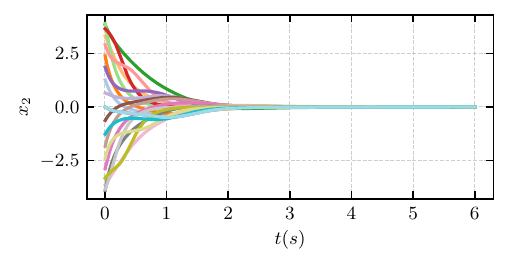}
  \caption{\label{fig:x2_SOP} $x_2$-$t$ profiles}
\end{subfigure}
\vfill
\begin{subfigure}{0.49\textwidth}
  \centering
  \includegraphics[scale=1.0]{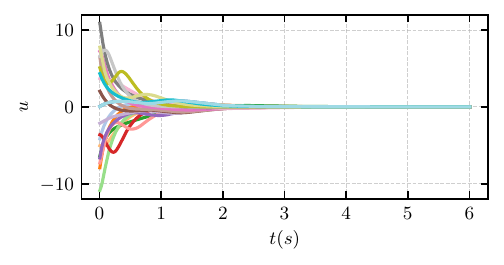}
  \caption{\label{fig:u_SOP} $u$-$t$ profiles}
\end{subfigure}
\caption{Profiles under the learned optimal control policy}
\label{fig:profiles_sop}
\end{figure}

The optimal trajectories are generated using Algorithm \ref{Algorithm2} and the data distribution is shown in Fig. \ref{fig:optimal_data_SOP}, where the solid red point represents equilibrium points, and black plus signs represent the desired initial states. The desired state space is set as $\mathcal D_d=\left\{ \boldsymbol{x}_0\in \mathbf{R}^n|\lVert \boldsymbol{x}_0 \rVert_2 \le 3.6 \right\} $. It can be seen that the generated optimal data covers the desired state space. In contrast, the standard BGOE method generates 100 optimal trajectories, which are shown in Fig. \ref{fig:optimal_data_SOP_NoSTM}. The generated optimal trajectories are not uniformly distributed and cannot cover the desired state space. Therefore, the proposed method is more efficient for generating desired optimal dataset.

Fig. \ref{fig:value_function_SOP} and Fig. \ref{fig:action_SOP} presents the approximate optimal value function and optimal control policy. Fig. \ref{fig:value_function_SOP_Compare} and Fig. \ref{fig:action_SOP_Compare} show the error between the learned value function and control policy and the analytical optimal solution, where the black dashed line indicates the boundary of the desired state space. It can be observed that the learned value function and control policy are very close to the analytical optimal solution within the desired state space. 

To further verify the effectiveness of the learned control policy, simulations are conducted at the boundary of the desired initial state space. The resulting state profiles are presented in Fig. \ref{fig:x1_SOP} and Fig. \ref{fig:x2_SOP}, while the corresponding control input profiles is shown in Fig. \ref{fig:u_SOP}. The system consistently converges to the equilibrium point from various initial states, demonstrating the validity of the learned optimal control policy.

\subsection{Winged-Cone Cruise Control}

\begin{table}[htbp]
  \centering
  \caption{Parameters for altitude control simulation}\label{tab:parameters_wcc}
  \begin{tabular}{lll} 
    \toprule
    {Parameters} & {Variables} & {Values} \\
    \midrule
    mass & $m$    & 9375 slugs \\
    radius of the Earth & $R_E$  & 20,903,500 ft \\
    reference area & $S$    & $3603\ \text{ft}^2$ \\
    gravitational constant & $\mu$    & $1.39\times10^{16}\ \text{ft}^3/\text{s}^2$ \\
    target height & $h_d$  & 110,000 ft \\
    cruising speed & $V$    & 15,060 ft/s \\
    angle of attack at trim condition & $\alpha_0$    & 0.0315 rad \\
    maximum angle of attack & $a_{\max}$    & $\pm$ 0.0872 rad (5 deg) \\
    \bottomrule
  \end{tabular}
\end{table}

The second example is a height control problem of simplified Winged-Cone model \cite{wangRobustNonlinearControl2000}. The model parameters are presented in Table \ref{tab:parameters_wcc}. The angle of attack is considered as control input, and the longitudinal dynamics can be expressed as
\begin{equation}
  \label{height system}
\left\{ \begin{aligned}
	\dot{h}&=V\sin \gamma\\
	\ddot{h}&=\frac{\left( L+T\sin \alpha \right) \cos \gamma}{m}-\frac{\left[ \left( \mu -V^2r \right) \cos ^2\gamma \right]}{r^2}\\
\end{aligned} \right. 
\end{equation}

The lift and drag force can be calculated by 
\begin{equation}
  \label{aerodynamics}
\begin{aligned}
	&L=\frac{1}{2}\rho V^2SC_L\\
	&D=\frac{1}{2}\rho V^2SC_D\\
\end{aligned}
\end{equation}
where lift coefficient $C_L$, drag coefficient $C_D$, air density $\rho$ and sound speed $a$ is expressed by empirical formulas as follows
\begin{equation}
  \label{aero coefficients}
\begin{aligned}
	&C_L=\alpha \left( 0.493+1.91/M \right)\\
	&C_D=0.0082\times \left( 171\alpha ^2+1.15\alpha +1 \right) \times \left( 0.0012M^2-0.054M+1 \right)\\
	&\rho =0.00238e^{-h/24000}\\
	&a=8.99\times 10^{-9}h^2-9.16\times 10^{-4}h+996\\
\end{aligned}
\end{equation}

\begin{figure*}[hbt!]
  \centering
  \includegraphics[scale=1.0]{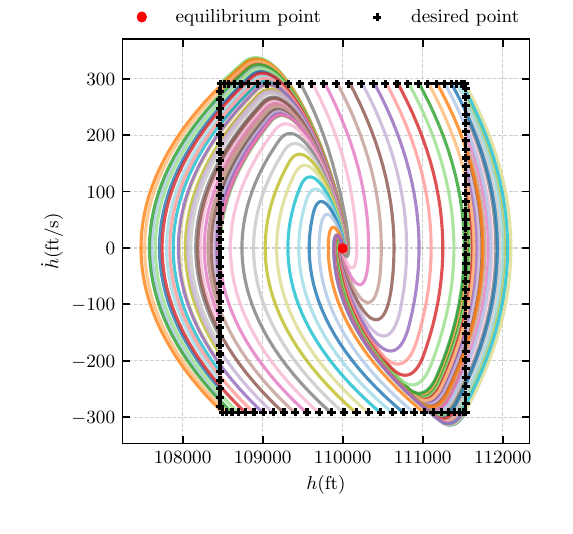}
  \caption{\label{fig:optimal_data_WCC} Optimal data in desired state space}
\end{figure*}

\begin{figure}[!htb]
  \centering
\begin{subfigure}{0.49\textwidth}
  \centering
  \includegraphics[scale=1.0]{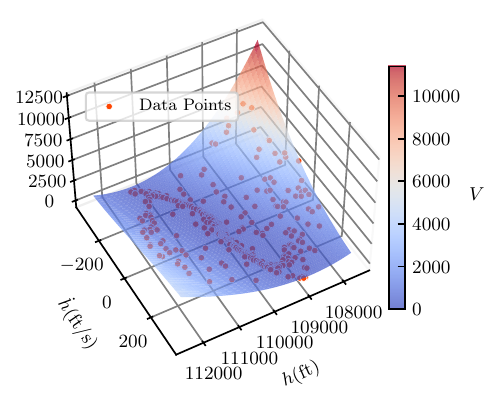}
  \caption{\label{fig:value_function_WCC} Optimal value function approximation}
\end{subfigure}
\hfill
\begin{subfigure}{0.49\textwidth}
  \centering
  \includegraphics[scale=1.0]{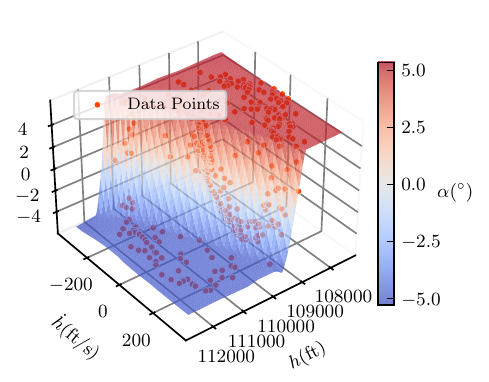}
  \caption{\label{fig:action_WCC} Optimal policy approximation}
\end{subfigure}
\caption{The learned optimal value function and optimal control policy}
\label{fig:optimal_value_policy_wcc}
\end{figure}

To reduce system complexity and facilitate the implementation of Algorithm \ref{Algorithm2}, we assume that the drag force is compensated by the thrust, which means the aircraft maintains constant speed. Let $x_1=h$ and $x_2=\dot{h}$, by substituting (\ref{aero coefficients}) and (\ref{aerodynamics}) into (\ref{height system}) and neglecting terms with small coefficients, the simplified longitudinal dynamics is derived as
\begin{equation}
\left\{ \begin{aligned}
	\dot{x}_1&=x_2\\
	\dot{x}_2&=64345.28\exp \left( -\frac{x_1}{24000} \right) \sqrt{1-\left( \frac{x_2}{15060} \right) ^2}\times \alpha -20.69\times\left[ 1-\left( \frac{x_2}{15060} \right) ^2 \right]\\
\end{aligned} \right. 
\end{equation}

The performance index is set as
\begin{equation}
J=\int_0^{\infty}{\left( \boldsymbol{x}-\boldsymbol{x}_e \right) ^T\boldsymbol{Q}\left( \boldsymbol{x}-\boldsymbol{x}_e \right) +R\left( \alpha -\alpha _0 \right) ^2\text{d}\tau}
\end{equation}
where $\boldsymbol{x}_e=\left[ h_d,0 \right] ^T=\left[ 110000,0 \right] ^T,\boldsymbol{Q}=\text{diag}\left( \left[ 0.0001,0.0001 \right] \right) ,R=1000$.

\begin{figure}[!htb]
  \centering
\begin{subfigure}{0.49\textwidth}
  \centering
  \includegraphics[scale=1.0]{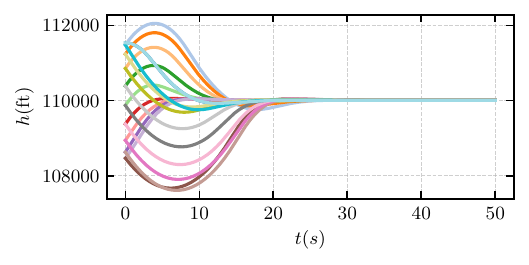}
  \caption{\label{fig:x1_WCC} $h$-$t$ profiles}
\end{subfigure}
\vfill
\begin{subfigure}{0.49\textwidth}
  \centering
  \includegraphics[scale=1.0]{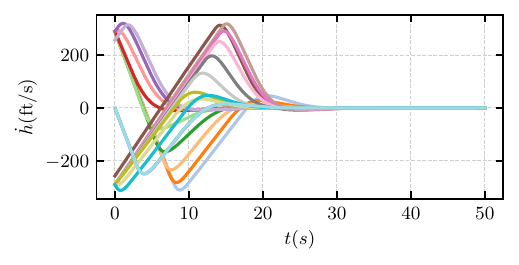}
  \caption{\label{fig:x2_WCC} $\dot h$-$t$ profiles}
\end{subfigure}
\vfill
\begin{subfigure}{0.49\textwidth}
  \centering
  \includegraphics[scale=1.0]{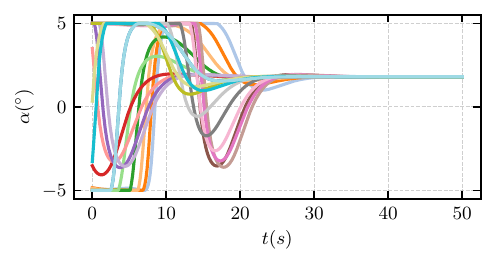}
  \caption{\label{fig:u_WCC} $\alpha$-$t$ profiles}
\end{subfigure}
\caption{Profiles under the learned optimal control policy}
\label{fig:profiles_wcc}
\end{figure}

The desired initial state space is set as $\mathcal D_d = \left\{ \boldsymbol{x}_0\in \mathbf{R}^n|108500\text{ ft}\le h\le 111500\text{ ft},-290\text{ ft/s}\le \dot{h}\le 290\text{ ft/s} \right\} $. Fig. \ref{fig:optimal_data_WCC} presents the distribution of the generated optimal trajectories, where the solid red point represents equilibrium points, and black plus signs represent the desired initial states. While some trajectories are out of the desired space, it's caused by the inherent dynamics of the system, and the desired initial state are all accurately generated. Fig. \ref{fig:value_function_WCC} and Fig. \ref{fig:action_WCC} presents the approximate optimal value function and optimal control policy. It can be seen that the approximate value function is positive definite and the control policy exhibits bang-bang characteristics. Simulations on different initial state are conducted, the state profiles and control profiles are presented in Fig. \ref{fig:x1_WCC}, Fig. \ref{fig:x2_WCC} and Fig. \ref{fig:u_WCC}. The aircraft successfully converges to the specified altitude from different initial states while complying with angle-of-attack constraints.

\subsection{Attitude Stabilization Control}
The third example is a rigid body attitude stabilization control problem. The kinematic and dynamic equations modeled by Euler angles can be expressed as
\begin{equation}
\left[ \begin{array}{c}
	\dot{\phi}\\
	\dot{\theta}\\
	\dot{\psi}\\
\end{array} \right] =\left[ \begin{matrix}
	1&		\tan \theta \sin \phi&		\tan \theta \cos \phi\\
	0&		\cos \phi&		-\sin \phi\\
	0&		\sin \phi /\cos \theta&		\cos \phi /\cos \theta\\
\end{matrix} \right] \left[ \begin{array}{c}
	p\\
	q\\
	r\\
\end{array} \right] 
\end{equation}
\begin{equation}
\left\{ \begin{aligned}
	\dot{p}&=\frac{1}{I_{xx}}\left[ \tau _x+qr\left( I_{yy}-I_{zz} \right) \right]\\
	\dot{q}&=\frac{1}{I_{yy}}\left[ \tau _y+pr\left( I_{zz}-I_{xx} \right) \right]\\
	\dot{r}&=\frac{1}{I_{zz}}\left[ \tau _z+pq\left( I_{xx}-I_{yy} \right) \right]\\
\end{aligned} \right. 
\end{equation}

\begin{figure*}[hbt!]
  \centering
  \includegraphics[scale=1.0]{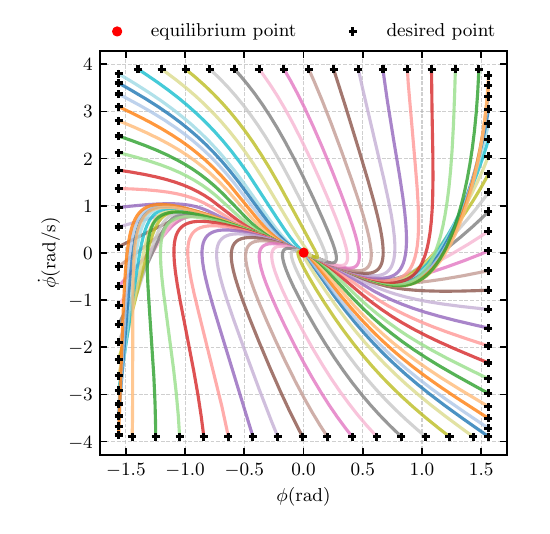}
  \caption{\label{fig:optimal_data_ATT} Optimal data in desired state space}
\end{figure*}

Set inertia parameters of the rigid body as $I_{xx}=0.0025\text{ kg} \cdot \text{m}^2, I_{yy}=0.0025\text{ kg} \cdot \text{m}^2, I_{zz}=0.0035\text{ kg} \cdot \text{m}^2$ and let $\boldsymbol{x}=\left( \phi ,\theta ,\psi ,p,q,r \right) ,\boldsymbol{u}=\left( \tau _x,\tau _y,\tau _z \right)$. The performance index is set as
\begin{equation}
J=\int_0^{\infty}{\boldsymbol{x}^T\boldsymbol{Qx}+\boldsymbol{u}^T\boldsymbol{Ru}\text{d}\tau}
\end{equation}
where $\boldsymbol{Q}$ are identity matrices and $\boldsymbol{R}=\text{diag}\left( \left[ 10^4,10^4,10^4 \right] \right)$.

\begin{figure}[!htb]
  \centering
\begin{subfigure}{0.49\textwidth}
  \centering
  \includegraphics[scale=1.0]{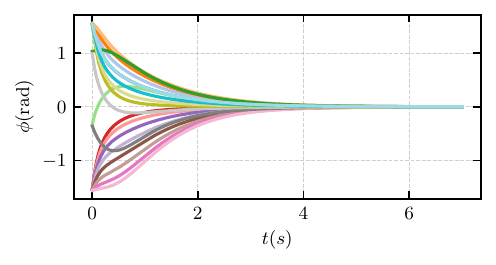}
  \caption{\label{fig:phi_ATT} $\phi$-$t$ profile}
\end{subfigure}
\hfill
\begin{subfigure}{0.49\textwidth}
  \centering
  \includegraphics[scale=1.0]{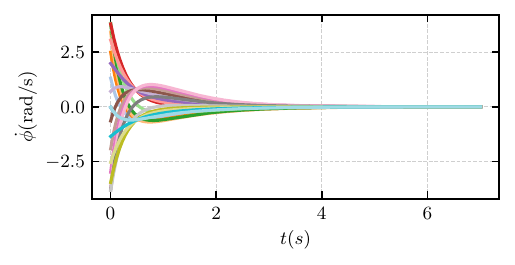}
  \caption{\label{fig:phi_dot_ATT} $\dot \phi$-$t$ profile}
\end{subfigure}
\vfill
\begin{subfigure}{0.49\textwidth}
  \centering
  \includegraphics[scale=1.0]{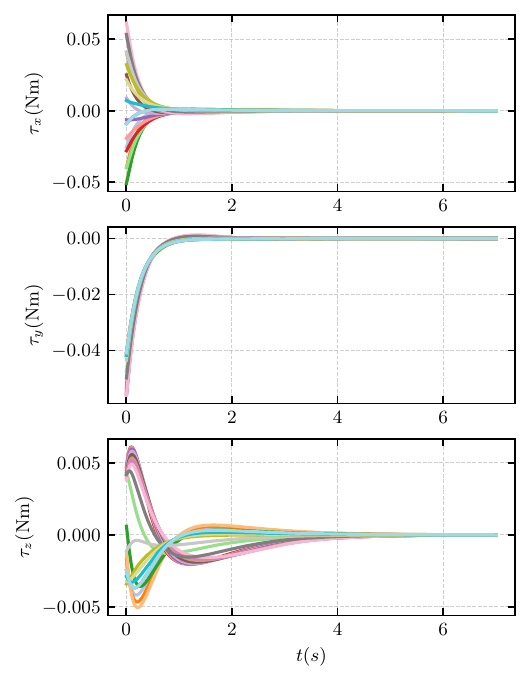}
  \caption{\label{fig:u_ATT} $\boldsymbol{u}$-$t$ profile}
\end{subfigure}
\caption{Profiles under the learned optimal control policy}
\label{fig:profiles_att}
\end{figure}

\begin{figure}[!htb]
  \centering
\begin{subfigure}{1\textwidth}
  \centering
  \includegraphics[scale=0.87]{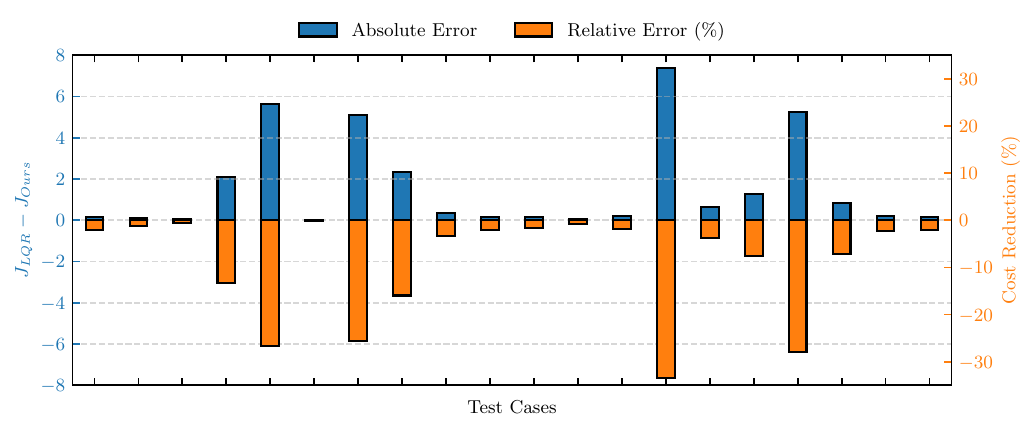}
  \caption{\label{fig:comparison_SOP} Example 1}
\end{subfigure}
\vfill
\begin{subfigure}{1\textwidth}
  \centering
  \includegraphics[scale=0.87]{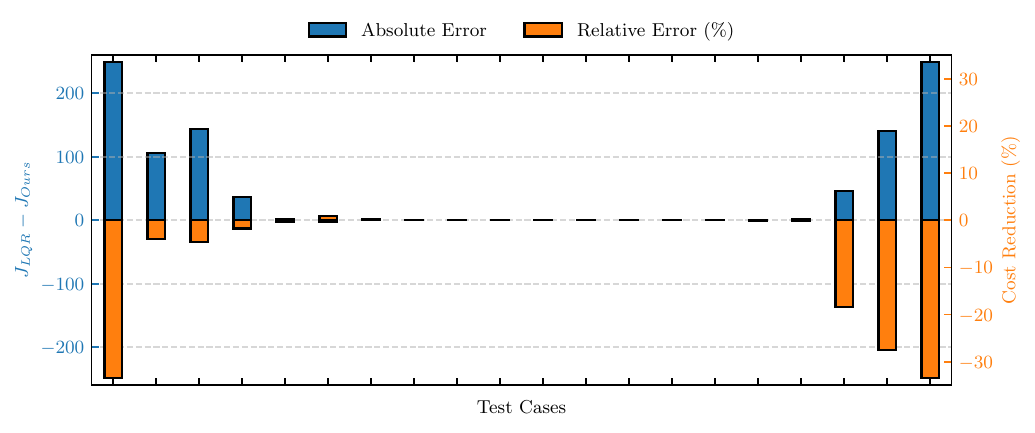}
  \caption{\label{fig:comparison_WCC} Example 2}
\end{subfigure}
\vfill
\begin{subfigure}{1\textwidth}
  \centering
  \includegraphics[scale=0.87]{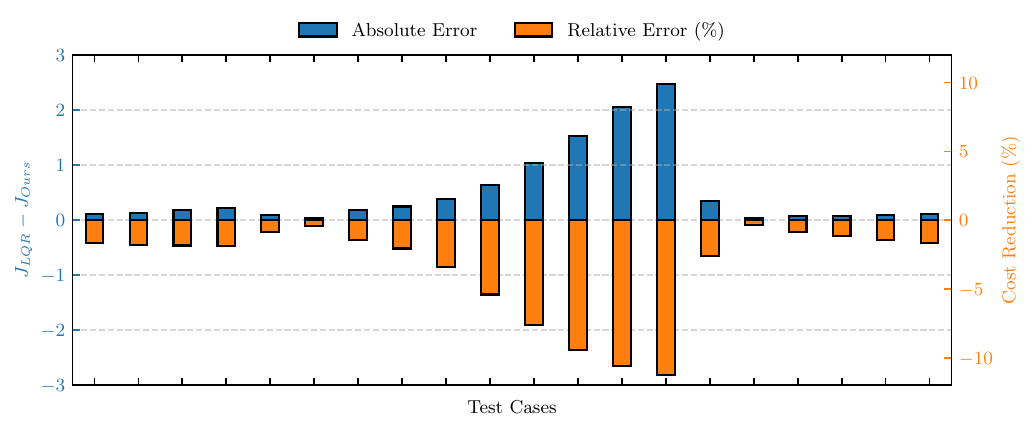}
  \caption{\label{fig:comparison_ATT} Example 3}
\end{subfigure}
\caption{Performance comparisons with LQR}
\label{fig:comparison_with_LQR}
\end{figure}

To facilitate visualization and reduce data generation volumes, data generation is restricted to the roll channel with varying initial states. The roll angle and rate are bounded at $-1.56\ \text{rad}\le \phi \le 1.56\ \text{rad}$, $-3.9\ \text{rad/s}\le \dot\phi \le 3.9\ \text{rad/s}$, respectively, while holding other initial states constant. The distribution of generated optimal trajectories is illustrated in Fig. \ref{fig:optimal_data_ATT}. State profiles under different initial states leveraging the learned optimal control policy are presented in Fig. \ref{fig:phi_ATT} and Fig. \ref{fig:phi_dot_ATT}, with corresponding control inputs shown in Fig. \ref{fig:u_ATT}. The approximated optimal control policy stabilizes the system in the desired state space, demonstrating the effectiveness of the proposed method.




To further evaluate the optimality of the learned control policy, we compare the performance index of the proposed method with LQR controllers in these three examples. In each example, 20 test cases are uniformly sampled at the boundary of the desired state space. Performance comparisons for the three examples are presented in Fig. \ref{fig:comparison_SOP}, Fig. \ref{fig:comparison_WCC} and Fig. \ref{fig:comparison_ATT}, respectively. To simultaneously exhibit both absolute and relative performance comparisons in one figure, the blue bars indicate absolute performance improvement, and the orange bars represent relative cost reduction percentages. When the blue bar is above the orange bar, it indicates an improvement in performance. It can be observed that the proposed method outperforms LQR controllers in almost all test cases of the three examples. Only in several test cases of the second example, the proposed method performs slightly worse than LQR controllers, which is caused by the approximation error of the neural networks. However, significant performance improvements are achieved in other test cases. In conclusion, comparative results across three examples demonstrates the optimality of the proposed framework for nonlinear optimal regulation problems.

\section{Conclusion}
\label{sec:Conclusion}
This paper proposes a learning-based framework to learn a stable optimal controller for infinite-time nonlinear optimal regulation problems. By leveraging equivalence between the HJB equation and the PMP condition, the standard BGOE method is extended to solve the infinite-time optimal regulation problem. Utilizing the state transition matrix of Hamiltonian systems, a state-transition-matrix-guided optimal dataset generation method is designed, enabling rapid generation of optimal datasets that effectively cover the desired state space. Compared to the standard BGOE method, the proposed method generates more complete optimal datasets and achieves higher data generation efficiency. Furthermore, by incorporating Lyapunov stability conditions, a stability-guaranteed learning framework is designed, capable of learning stable approximate optimal control policies. Simulations conducted on three examples demonstrate that the proposed method can efficiently generate optimal datasets and learn stable near-optimal control policies, indicating significant performance improvement compared to traditional LQR control. The current method still faces challenges related to the curse of dimensionality in dataset generation for higher-dimensional scenarios. Future studies will explore more efficient data generation and policy learning techniques to further enhance solution efficiency.
\bibliography{references}

\end{document}